\documentclass[twocolumn,showpacs,preprintnumbers,amsmath,amssymb,superscriptaddress,aps]{revtex4}

\usepackage{graphicx}
\usepackage{dcolumn}
\usepackage{bm}
\usepackage{color} 
\usepackage{ulem}

\begin{document}

\title{Direct observation of heterogeneous valence state in Yb-based quasicrystalline approximants}

\author{M.~Matsunami}
 \altaffiliation[Electronic address: ]{matunami@toyota-ti.ac.jp}
\affiliation{Toyota Technological Institute, Nagoya 468-8511, Japan}
\affiliation{RIKEN SPring-8 Center, Sayo, Hyogo 679-5148, Japan}
\author{M.~Oura}
\author{K.~Tamasaku}
\author{T.~Ishikawa}
\affiliation{RIKEN SPring-8 Center, Sayo, Hyogo 679-5148, Japan}
\author{S.~Ideta}
\author{K.~Tanaka}
\affiliation{UVSOR Facility, Institute for Molecular Science, Okazaki 444-8585, Japan}
\author{T.~Takeuchi}
 \affiliation{Toyota Technological Institute, Nagoya 468-8511, Japan}
\author{T.~Yamada}
\author{A.~P.~Tsai}
\affiliation{Institute of Multidisciplinary Research for Advanced Materials, Tohoku University, Sendai, Japan}
\author{K.~Imura}
\author{K.~Deguchi}
\author{N.~K. Sato}
\affiliation{Department of Physics, Graduate School of Science, Nagoya University, Nagoya 464-8602, Japan}
\author{T.~Ishimasa}
\affiliation{Division of Applied Physics, Graduate School of Engineering, Hokkaido University, Sapporo 060-8628, Japan}

\date{\today}

\begin{abstract} 
We study the electronic structure of Tsai-type cluster-based quasicrystalline approximants, Au$_{64}$Ge$_{22}$Yb$_{14}$ (AGY-I), Au$_{63.5}$Ge$_{20.5}$Yb$_{16}$ (AGY-II), and Zn$_{85.4}$Yb$_{14.6}$ (Zn-Yb), by means of photoemission spectroscopy. 
In the valence band hard x-ray photoemission spectra of AGY-II and Zn-Yb, we separately observe a fully occupied Yb\,4$f$ state and a valence fluctuation derived Kondo resonance peak, reflecting two inequivalent Yb sites, a single Yb atom in the cluster center and its surrounding Yb icosahedron, respectively. 
The fully occupied 4$f$ signal is absent in AGY-I containing no Yb atom in the cluster center. 
The results provide direct evidence for a heterogeneous valence state in AGY-II and Zn-Yb. 
\end{abstract}


\maketitle

Mixed valence (MV) phenomena have been one of the central issues of condensed matter physics, particularly in the field of heavy fermion systems \cite{MV_Varma}. 
The mixing of different electronic configurations, e.g., nonmagnetic Yb$^{2+}$ (4$f^{14}$) and magnetic Yb$^{3+}$ (4$f^{13}$), in the quantum mechanical ground state can be responsible for a wide variety of strongly correlated electronic properties. 
The unconventional quantum criticality observed in Yb-based compounds such as YbRh$_2$Si$_2$ \cite{YbRh2Si2_Trovarelli} and $\beta$-YbAlB$_4$ \cite{YbAlB4_Nakatsuji} has been investigated in association with the MV state of Yb ions \cite{Watanabe}. 
Recently, the quantum criticality in the Yb-based quasicrystal Au-Al-Yb exhibiting MV \cite{Watanuki_AAY} has attracted special attention as a specific manifestation of critical wave functions predicted in quasiperiodic systems \cite{Deguchi_NM,Shaginyan,Matsukawa,Andrade}. 
The term ``MV'' here means a dynamically fluctuating one and frequently is referred to as ``valence fluctuation.'' 
The concept of valence fluctuation is completely different from the spatially heterogeneous valence state, such as static charge separation or ordering. 
Nonetheless, for systems showing an intermediate valence, it is not easy to directly distinguish between valence fluctuations and spatial valence heterogeneity by means of core-level spectroscopies including photoemission, absorption, and inelastic scattering. 
Since valence fluctuations are caused by the hybridization between 4$f$ and conduction bands, we have to directly observe the electronic structure in the vicinity of the Fermi level ($E_{\rm F}$), e.g., Kondo resonance peak.

As an example that distinguishing between valence fluctuations and spatial valence heterogeneity is highly required, here we consider a Tsai-type cluster consisting of successive concentric shells, which is one of the basic structural units in quasicrystals \cite{Tsai_CdYb}. 
In the case of Yb-based quasicrystals or approximants, as shown in Fig.~1(a), the shells of an orientationally disordered non-Yb tetrahedron, non-Yb dodecahedron, Yb icosahedron, and non-Yb icosidodecahedron are successively arranged from the inside to the outside. 
The Au-Al-Yb quasicrystal and approximant described above possess this type-I Tsai cluster \cite{Ishimasa_AAY}. 
On the other hand, the central tetrahedron is occasionally replaced by a single Yb atom (type-II), as shown in Fig.~1(b). 
Such a central Yb atom (labeled as Yb2) is crystallographically inequivalent to that (Yb1) of the vertex of an icosahedron. 
Au-Ge-Yb (AGY) systems are known to have two types of compounds consisting of these two types of Tsai clusters \cite{Lin_AGY}. 
Hereafter, AGY compounds with type-I and type-II Tsai clusters are referred to as AGY-I and AGY-II, respectively \cite{Deguchi_AGY_JPSJ}. 
As another remarkable case, these two types of Tsai clusters coexist in a single compound of the Zn-Yb system \cite{Fornasini_ZnYb}, which is denoted as Zn$_{17}$Yb$_{3}$ \cite{Bruzzone_ZnYb} or Zn$_{6}$Yb \cite{Fornasini_ZnYb}. 
More recently, in Au-Si-Tb systems, it has been reported that a variation from a type-I to type-II Tsai cluster can be continuously controlled by changing the composition \cite{Tb-Au-Si}. 
All these systems belong to the 1/1 approximant of a Tsai-type icosahedral quasicrystal.

In this Rapid Communication, we approach the local electronic states depending on the types of Tsai clusters in the quasicrystalline approximants AGY-I, AGY-II, and Zn-Yb using photoemission spectroscopy (PES). 
In the valence band hard x-ray PES (HAXPES) spectra of AGY-II and Zn-Yb, we separately observe two components of the Yb$^{2+}$\,4$f_{7/2}$ state, one at/near $E_{\rm F}$ and the other well away from $E_{\rm F}$. 
The latter is absent in AGY-I. 
We demonstrate that Yb valence heterogeneity occurs between the valence fluctuation state at the Yb1 site and the purely divalent state at the Yb2 site.

\begin{figure}[t]
\begin{center}
\includegraphics[width=0.46\textwidth]{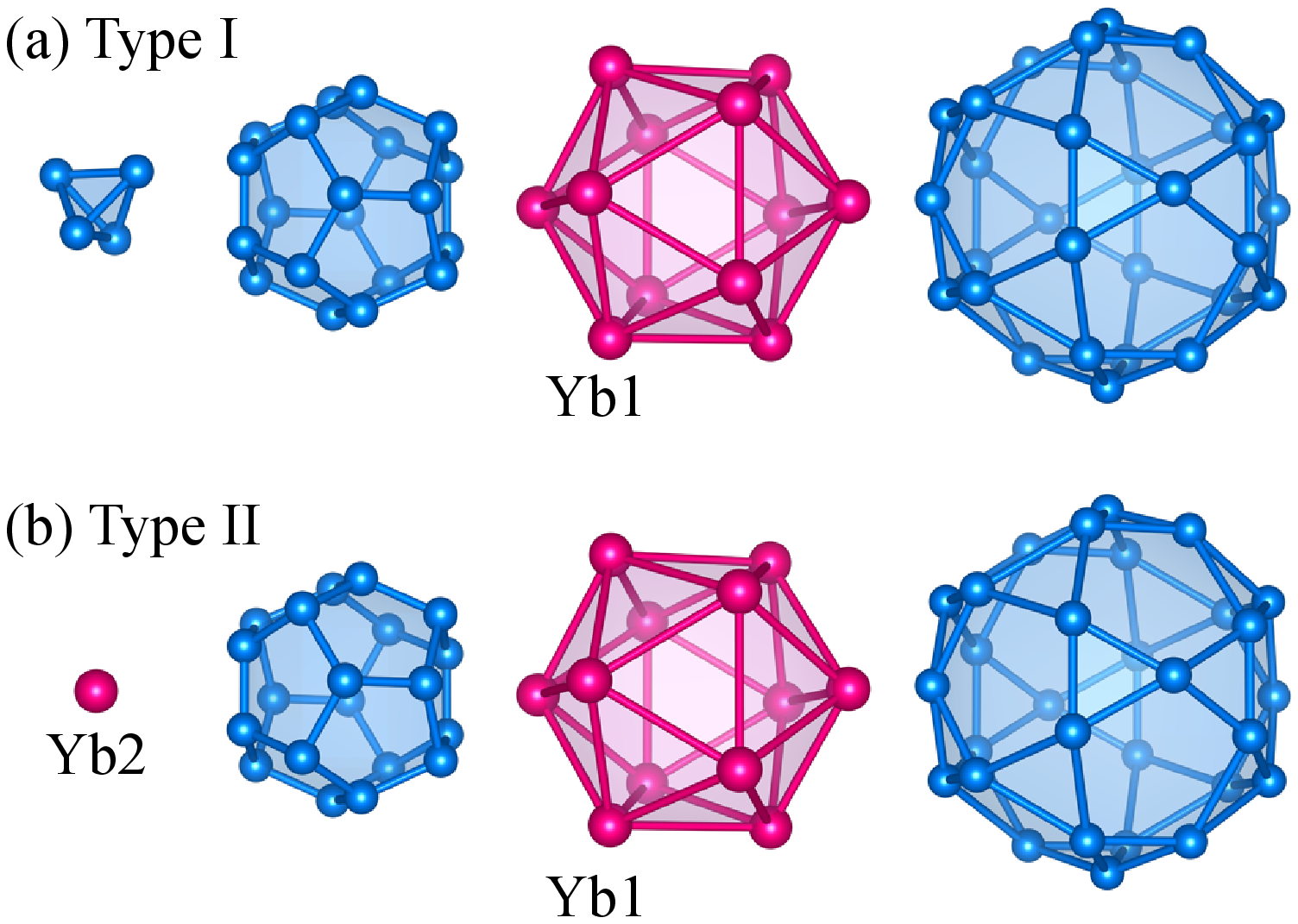}
\caption{
Two types of successive shell structures of the Tsai clusters. 
For Yb-based quasicrystalline approximants in this study, the pink spheres correspond to Yb atoms and the blue spheres are non-Yb atoms such as Au, Ge, or Zn atoms \cite{Lin_AGY,Fornasini_ZnYb,Bruzzone_ZnYb,Deguchi_AGY_JPSJ}. 
The difference is of the innermost component; (a) non-Yb tetrahedron or (b) single Yb atom. 
The former is orientationally disordered in most cases. 
Successively surrounding non-Yb dodecahedron, Yb icosahedron, and non-Yb icosidodecahedron are common to both clusters. 
The inequivalent Yb atoms are labeled as Yb1 for the vertex of the icosahedron and Yb2 for the center of the cluster. 
} 
\end{center}
\end{figure}

\begin{figure}[t]
\begin{center}
\includegraphics[width=0.46\textwidth]{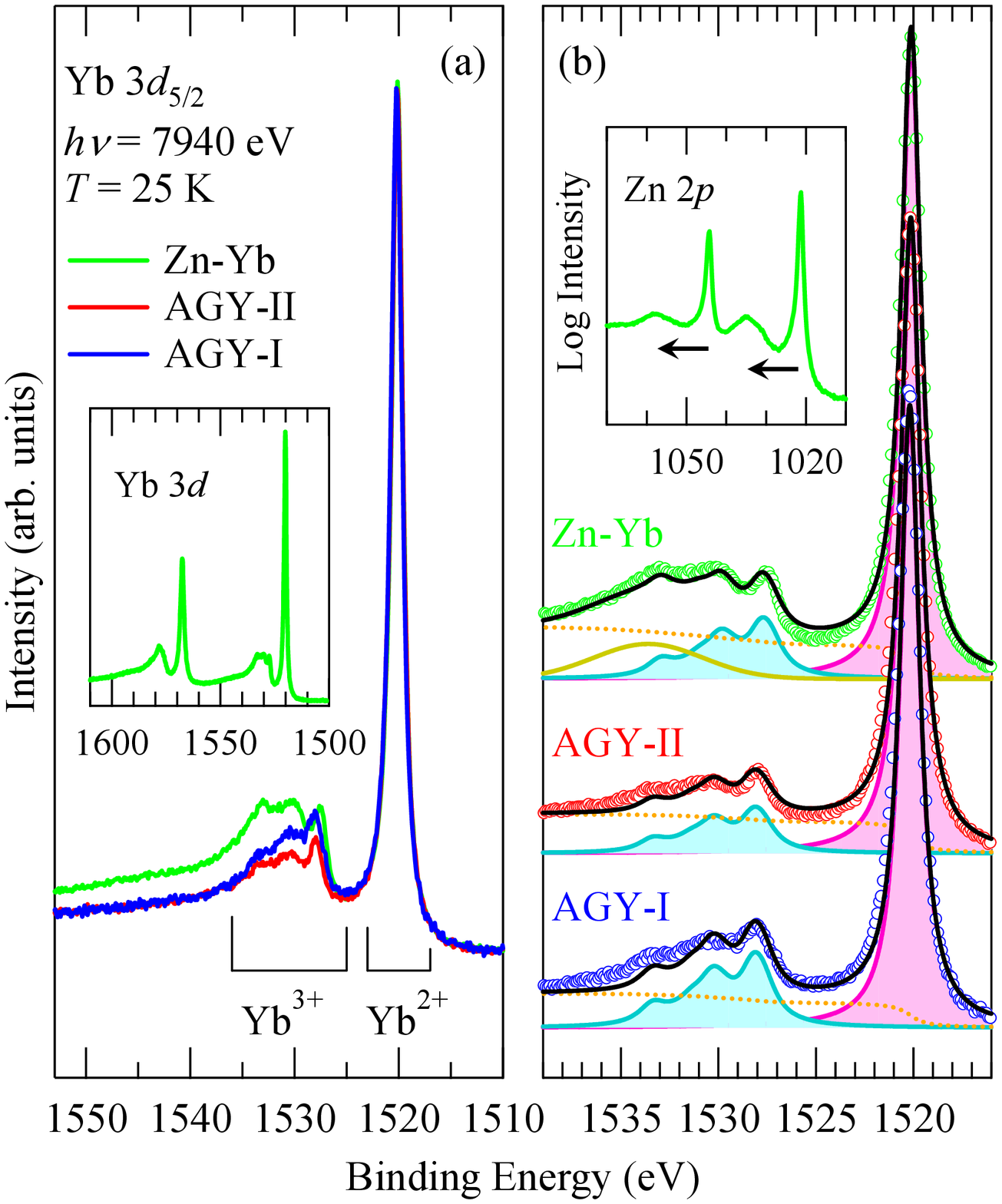}
\caption{
(a) Yb\,3$d_{5/2}$ core-level HAXPES spectra of AGY-I, AGY-II, and Zn-Yb. 
The inset shows the overall 3$d$ spectrum of Zn-Yb. 
(b) The fitting results (black line), using the atomic multiplet model for free Yb$^{2+}$ (pink shading) and Yb$^{3+}$ (cyan shading) ions, and the integral-type background (orange dotted line).
The inset shows the Zn\,2$p$ core-level spectrum of Zn-Yb in a logarithmic scale plot. 
Arrows represent the plasmon energy, which is reflected as the yellow line in fitting of the Yb\,3$d_{5/2}$ spectrum of Zn-Yb. 
} 
\end{center}
\end{figure}

Polycrystalline samples of AGY-I, AGY-II and Zn-Yb were prepared. 
The details of the preparation and characterization of AGY-I and AGY-II have been reported elsewhere \cite{Deguchi_AGY_JPSJ}. 
Their composition was Au$_{64}$Ge$_{22}$Yb$_{14}$ and Au$_{63.5}$Ge$_{20.5}$Yb$_{16}$, respectively. 
The sample of Zn-Yb was grown by a method similar to the one reported previously \cite{Bruzzone_ZnYb}. 
The composition was confirmed as Zn$_{85.4}$Yb$_{14.6}$ and the ratios for the two types of Tsai clusters (Fig.~1) were 1/3 for type-I and 2/3 for type-II. 
Synchrotron-radiation-based angle-integrated \cite{BZ} PES was carried out at three undulator beamlines with respective $h\nu$'s. 
The HAXPES ($h\nu$ $=$ 7940\,eV) was performed at BL19LXU in SPring-8 with a Scienta R4000-10kV electron analyzer. 
The soft x-ray (SX-) PES ($h\nu$ $=$ 600\,eV) was performed at BL17SU in SPring-8 with a Scienta SES-2002. 
The VUV-PES ($h\nu$ $=$ 20\,eV) was performed at BL7U of UVSOR at the Institute for Molecular Science with an MBS A-1. 
The energy resolution was set to 250\,meV for the core-level HAXPES, 100\,meV for the valence band HAXPES and SX-PES, and 10\,meV for VUV-PES. 
The temperature was set to the lowest value for each apparatus: 25\,K for HAXPES, 20\,K for SX-PES, and 12\,K for VUV-PES. 
Clean sample surfaces were obtained by fracturing the samples $in$ $situ$. 
The binding energy ($E_{\rm B}$) of the samples was calibrated using $E_{\rm F}$ of the evaporated Au films.

Figure~2(a) shows the Yb\,3$d_{5/2}$ core-level spectra of AGY-I, AGY-II, and Zn-Yb, together with the overall 3$d$ spectrum of Zn-Yb in the inset. 
For all samples, both the Yb$^{2+}$ peak and Yb$^{3+}$ multiplet are confirmed, indicating their MV state. 
As described above, however, we cannot conclude valence fluctuations only from the core-level spectra. 
To quantify their Yb mean valence, a spectral fitting was conducted using a standard procedure \cite{YbAl2_HAXPES}, as shown in Fig. 2(b). 
The Yb mean valence is evaluated as +2.26, +2.18, and +2.22 for AGY-I, AGY-II, and Zn-Yb, respectively. 
The energy loss satellite due to plasmon excitations is not observed in AGY-I and AGY-II. 
On the other hand, as is seen for the Zn\,2$p$ spectrum in the inset of Fig.~2(b), the plasmon feature is clearly observed in Zn-Yb.

Figure 3 shows the valence band HAXPES spectra of AGY-I, AGY-II, and Zn-Yb. 
The two main peaks at $\sim$0.2 and 1.5\,eV are derived from the Yb$^{2+}$ 4$f$ doublet ($J$ $=$ 7/2 and 5/2, respectively) corresponding to the 4$f^{13}$ final state. 
The 4$f_{7/2}$ peak certainly crosses $E_{\rm F}$, providing direct evidence of valence fluctuations, as discussed later in detail. 
Note that, for AGY-II and Zn-Yb, additional features are observed at $\sim$0.9 and 2.2\,eV, as indicated by the arrows in Fig.~3. 
The same energy separation as the main peaks suggests that the feature is also associated with the Yb$^{2+}$\,4$f$ states. 
Considering their smaller intensity and their absence only in AGY-I, the feature can be derived from the Yb2 site in Fig.~1. 
Since the feature locates well away from $E_{\rm F}$, the Yb ions at the Yb2 site can be identified as purely divalent ones \cite{Yb2+_Matsunami}. 
Alternatively, the main peaks are attributed to the Yb1 site. 
The position of the Yb\,4$f_{7/2}$ main peak, which is closely related to the Kondo temperature as a measure of hybridization strength with the conduction band, of AGY-I, AGY-II, and Zn-Yb is relatively away from $E_{\rm F}$ ($E_{\rm B}$$\sim$0.2\,eV), consistent with their Yb mean valence being closer to divalent. 
For YbAl$_2$ possessing a similar Yb valence of +2.2 \cite{YbAl2_HAXPES}, the 4$f_{7/2}$ peak also locates at a similar $E_{\rm B}$ of 0.17\,eV and certainly contributes to the Fermi surfaces through the highly dispersive bands caused by the strong hybridization \cite{YbAl2_ARPES}. 
From this point of view, the 4$f_{7/2}$ main peak for AGY-I, AGY-II, and Zn-Yb can be also identified as a Kondo resonance peak responsible for valence fluctuations.

\begin{figure}[t]
\begin{center}
\includegraphics[width=0.35\textwidth]{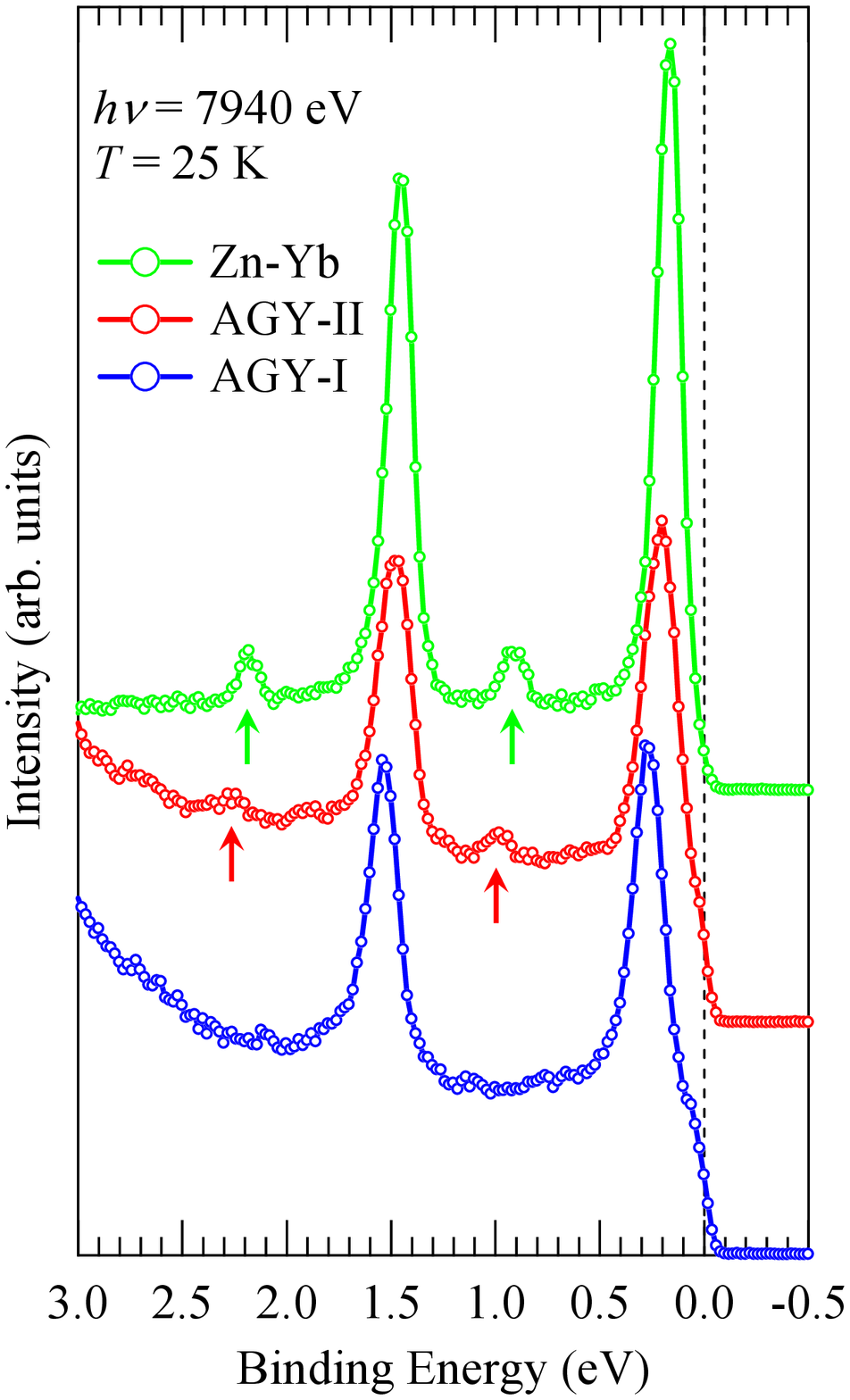}
\caption{
Valence band HAXPES spectra of AGY-I, AGY-II, and Zn-Yb. 
The spectra are normalized to an integrated intensity up to $E_{\rm B}$ = 0.5\,eV.
Arrows indicate the additional feature discussed in the text. 
} 
\end{center}
\end{figure}

\begin{figure}[t]
\begin{center}
\includegraphics[width=0.46\textwidth]{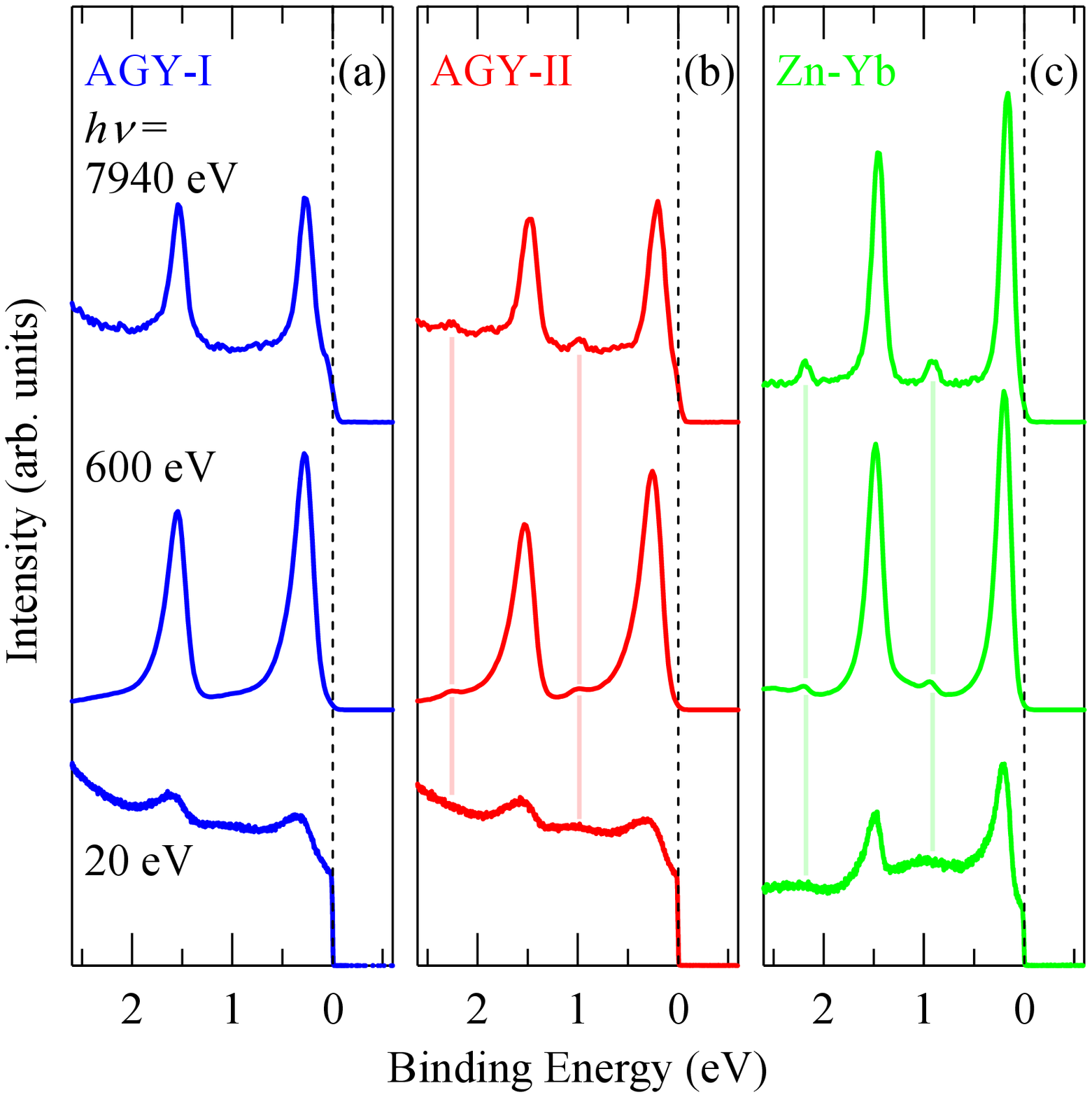}
\caption{
Valence band spectra of (a) AGY-I, (b) AGY-II, and (c) Zn-Yb measured with $h\nu$'s of hard x-ray (7940\,eV), SX (600\,eV), and VUV (20\,eV). 
The spectra are normalized to an integrated intensity up to $E_{\rm B}$ = 0.5\,eV.
} 
\end{center}
\end{figure}

Next, we compare the valence band spectra of AGY-I, AGY-II, and Zn-Yb measured with several $h\nu$'s, as shown in Fig.~4. 
For all samples, the SX-PES spectra exhibit the most prominent feature of the 4$f$ main peaks, reflecting the larger photoionization cross sections of the 4$f$ electrons. 
For VUV-PES spectra, these peaks are suppressed, particularly in AGY-I and AGY-II. 
Regarding the ``Yb2 peaks'' in AGY-II and Zn-Yb, we can see their rapid depletion with decreasing $h\nu$, i.e., increasing surface sensitivity in PES \cite{MEE}. 
The feature seems to be completely buried in the VUV-PES spectra. 
The position of the Yb2 peaks just overlaps with the surface component derived from purely divalent Yb surface atoms. 
As is well known in PES studies of the elemental Yb metal \cite{Yb_metal_Surface}, in which the Yb mean valence is purely divalent even in the bulk \cite{Yb2+_Matsunami}, the presence of surface Yb atoms with different coordination numbers gives rise to surface peaks at different binding energies, leading to a broader peak than that in the bulk. 
Therefore, the bulk and surface components of fully occupied 4$f$ states can be distinguished in terms of peak broadness. 
Indeed, as is clear for Zn-Yb in Fig.~4, we find the sharpness of Yb2 peaks in HAXPES spectra compared with the surface components in VUV spectra. 
As a result, the Yb2 peaks observed in the HAXPES spectra of AGY-II and Zn-Yb are derived from the bulk states, not the surface states. 
Importantly, the Yb2 peaks are clearly suppressed even in the SX-PES spectra. 
Considering a scale of the cluster (14.3-14.7\,{\AA}) \cite{Fornasini_ZnYb,Deguchi_AGY_JPSJ} comparable to the escape depth of photoelectrons (roughly 10-20\,{\AA}) \cite{Tanuma_IMFP} in SX-PES with $h\nu$ = 600\,eV, even SX-PES is insufficient for probing the cluster-derived electronic states in the bulk.

Our findings thus indicate that Yb valence heterogeneity in type-II Tsai cluster occurs between the valence fluctuation state at the Yb1 site and the purely divalent state at the Yb2 site. 
In this case, the mean valence at the Yb1 sites in AGY-II and Zn-Yb should be slightly larger than that estimated by the Yb\,3$d$ core-level spectra described above. 
It should be noted that two heterogeneous Yb$^{2+}$ components are not distinguished in the Yb\,3$d$ core-level spectra, probably due to their energy separation being smaller than the peak width. 
Such a heterogeneous valence state in the Yb\,4$f$ spectra reflecting the crystallographically inequivalent Yb sites has been so far inferred in Yb$_5$Si$_3$ using a multiple peak fitting \cite{Yb5Si3_PES}. 
On the other hand, the present results clearly reveal that the heterogeneous valence state can be directly and separately observed in the valence band spectra measured with high-resolution and bulk-sensitive HAXPES.

For Yb-based quasicrystals and approximants with Tsai-type clusters, the interatomic distance between neighboring Yb ions is relatively large:  For AGY-II (Zn-Yb), the distance between neighboring Yb1 atoms is 5.5\,{\AA} (5.3-5.4\,{\AA}) and that between Yb1 and Yb2 atoms is 5.3\,{\AA} (5.0\,{\AA}). 
Therefore, the Yb-Yb hybridization can be negligible, as also pointed out in the study of Au-Al-Yb systems \cite{Matsukawa}. 
Instead, the Yb\,4$f$ states can hybridize with the conduction electrons of the nearest-neighbor non-Yb atoms. 
Note that the distance of 3.2\,{\AA} in AGY-II (3.1\,{\AA} in Zn-Yb) between Yb1 and nearest neighbor non-Yb atoms is smaller than that of 3.7\,{\AA} (3.6\,{\AA}) between Yb2 and nearest-neighbor non-Yb atoms. 
Consequently, we expect that the 4$f$ electrons at the Yb1 site are likely to hybridize with the conduction electrons of the nearest-neighbor non-Yb atoms in the inner dodecahedron or outer icosidodecahedron (Fig.~1), but the 4$f$ shells at the Yb2 site are fully filled without any hybridization. 
This finding provides an important insight into the local electronic states in the Tsai-type clusters.

Finally, we briefly comment on the difference between AGY's and Zn-Yb. 
In Fig.~4, we observe a stronger degradation of 4$f$ peaks in AGY-I and AGY-II with decreasing $h\nu$ compared to that of Zn-Yb in spite of their similarity both for the Yb mean valence and Kondo temperature. 
One possible explanation is that an inhomogeneity of the hybridization target of 4$f$ electrons due to the mixed atomic sites of Au/Ge \cite{Lin_AGY} can make their Kondo state unstable, leading to peak broadening. 
Further study is necessary for clarifying such microscopic and local electronic interactions in association with the transport properties and the observability of plasmon excitations as described above.

In conclusion, we performed PES of the quasicrystalline approximants with two types of Tsai clusters, AGY-I, AGY-II, and Zn-Yb. 
The Yb mean valence is estimated to be +2.26, +2.18, and +2.22, respectively. 
In the valence band HAXPES spectra of AGY-II and Zn-Yb, we separately observe the Kondo resonance peak and the fully occupied Yb\,4$f$ states, which are derived from the inequivalent Yb1 and Yb2 sites, respectively. 
The latter is absent in AGY-I. 
The results reveal that Yb valence heterogeneity in type-II Tsai clusters occurs between the valence fluctuation state at the Yb1 site and the purely divalent state at the Yb2 site.

The HAXPES and SX-PES measurements were carried out with the approval of the RIKEN SPring-8 Center (Proposals No. 20150035 and No. 20150088, respectively). 
The VUV-PES was performed by the Use-of-UVSOR Facility Program of the Institute for Molecular Science (2015).


\begin{thebibliography}{99}
\bibitem{BZ}
In this work, the angle- or momentum-resolved effect is excluded, because the samples, in addition to being polycrystalline, possess large lattice parameters of more than 14\,{\AA} and hence the first Brillouin zone is sufficiently smaller than the momentum space covered even with $h\nu$ = 20\,eV.
\bibitem{MEE}
As another or additional origin for the $h\nu$ variation of the Yb2 peaks, the possible matrix element effect, which can be pronounced in the lower-$h\nu$ PES, also should be considered.
\end{thebibliography}

\end{document}